\documentclass[a4paper]{article}

\usepackage[T1]{fontenc}
\usepackage[utf8x]{inputenc}
\usepackage[english]{babel}
\usepackage[toc,page]{appendix}
\usepackage[colorlinks=true, allcolors=blue]{hyperref}

\newcommand{\email}[1]{\href{mailto:#1}{\tt{\nolinkurl{#1}}}}
\newcommand{\orcid}[1]{ORCID: \href{https://orcid.org/#1}{\tt{\nolinkurl{#1}}}}

\usepackage[parfill]{parskip}

\usepackage{fancyhdr}
\usepackage{natbib}
\usepackage{authblk}
\usepackage[a4paper,top=3cm,bottom=2cm,left=3cm,right=3cm,marginparwidth=1.75cm]{geometry}

\usepackage{amsmath}
\usepackage{graphicx}
\usepackage{booktabs}

\usepackage[colorinlistoftodos]{todonotes}

\title{Wide field-of-view flat lens: an analytical formalism}
\author[1*]{Fan Yang}
\author[2]{An Sensong}
\author[1]{Mikhail Y. Shalaginov}
\author[2]{Zhang Hualiang}
\author[1,3*]{Juejun Hu}
\author[1,3*]{Tian Gu}
\affil[1]{Department of Materials Science $\&$ Engineering, Massachusetts Institute of Technology, Cambridge, Massachusetts 02139, United States}
\affil[2]{Department of Electrical $\&$ Computer Engineering, University of Massachusetts Lowell, Lowell, Massachusetts 01854, United States}
\affil[3]{Materials Research Laboratory, Massachusetts Institute of Technology, Cambridge, Massachusetts 02139, United States}
\affil[*]{Corresponding author: \email{yangf@mit.edu, hujuejun@mit.edu, gutian@mit.edu}}
\date{}
\begin{document}
\maketitle
\thispagestyle{fancy}

\begin{abstract}
Wide field-of-view (FOV) optics are widely used in various imaging, display, and sensing applications. While conventional wide FOV optics rely on cascading multiple elements to suppress coma and other aberrations, it has recently been demonstrated that diffraction-limited, near-180\textdegree\ FOV operation can be achieved with a single-piece flat fisheye lens designed via iterative numerical optimization [Nano Lett. {\bfseries 20}, 7429(2020)]. Here we derive an analytical formalism to enable physics-informed and computationally efficient design of wide FOV flat lenses based on metasurfaces or diffractive optical elements (DOEs). Leveraging this analytical approach, we further quantified trade-offs between optical performance and design parameters in wide FOV metalenses.
\end{abstract}

\section{Introduction}

Wide field-of-view (WFOV) optics, exemplified by fisheye lenses featuring a FOV close to or even exceeding 180$^\circ$, are widely employed in landscape photography, security surveillance, meteorology, and image projection \cite{liu2018metasurface,jiang2014broadband,lee2018metasurface,guo2018high}. In recent years, they are also starting to gain traction in emerging electronics and optics products, enabling panoramic cameras and 3-D depth sensors, AR/VR optics rendering immersive experiences, omnidirectional computer vision systems, and new biomedical imaging instruments. To fulfill these application demands, suppression of off-axis optical aberrations such as coma, astigmatism and field curvature is crucial to realizing high-quality WFOV optics. The traditional approach for aberration mitigation entails cascading multiple lens elements, which however increases the size, weight, complexity, and cost of the optical system.

Flat optics based on optical metasurfaces or diffractive optical elements (DOEs) offer an alternative solution to expand the FOV of optical systems. One scheme involves stacking multiple metasurfaces, and diffraction-limited FOVs up to 56$^\circ$ have been attained using this method \cite{groever2017meta,arbabi2016miniature}. Combining a single-layer metasurface or diffractive lens with a physical or virtual optical aperture provides an architecturally simpler approach \cite{buralli1989design,shalaginov2019single,engelberg2020near,martins2020metalenses,liang2019high,lassalle2021imaging,chen2020chip}. In particular, a single-element fisheye metalens was demonstrated with > 170$^\circ$ diffraction-limited FOV \cite{shalaginov2020single}. This unprecedented performance was accomplished through iterative numerical optimization of the metasurface optical phase profile, a computationally intensive process precluding extensive exploration of the full design parameter space.

In this letter, we derive an analytical solution to the optimum phase profile of a WFOV flat lens assuming the single-layer geometry, yielding results in excellent agreement with numerically optimized designs but without requiring computationally intensive optimization. The analytical solution is generically applicable to different operation wavelength ranges, lens/substrate materials, and meta-atom or diffractive element designs. Finally, we derive an expression relating design parameters with focusing and imaging performance and investigate the design trade-offs in realizing WFOV flat lenses.

\section{Analycial formalism}

\begin{figure*}[htbp]
\centering
\includegraphics[width=\linewidth]{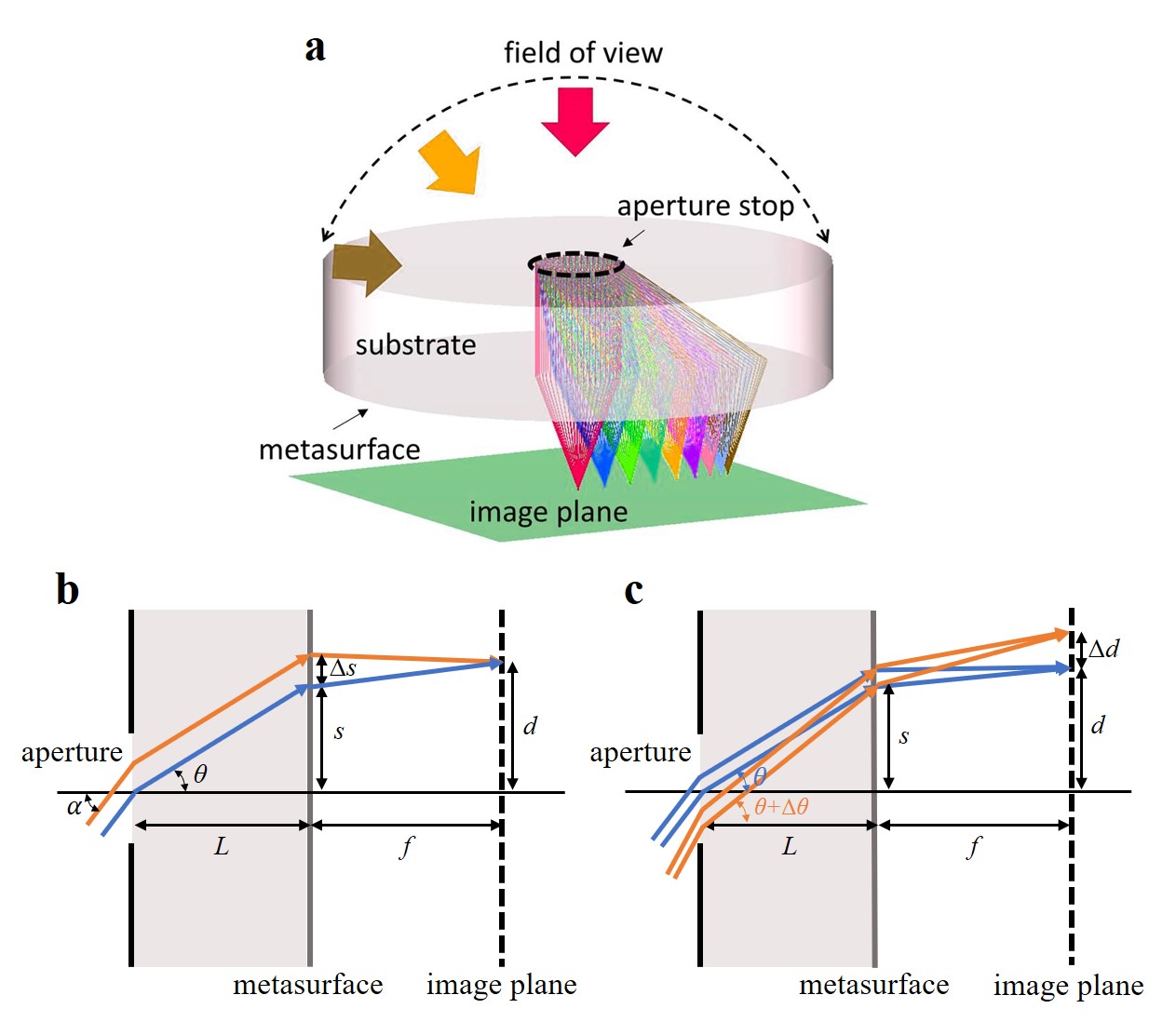}
\caption{Schematic illustration of WFOV metalens design. (a) 3-D structure. (b) Illustration of the phase profile derivation. (c) Illustration of the image height derivation. Fig. 1(a) is adapted from \cite{shalaginov2020single}.}
\label{fig:1}
\end{figure*}

The basic concept of the single-layer WFOV flat lens is illustrated in Fig. \ref{fig:1}a. An aperture is placed at the front surface of a substrate and a metasurface (or a DOE surface) is patterned on the back surface to act as an optical phase mask. Beams from different angles of incidence (AOIs) are refracted at the front surface and arrive at different and yet continuous portions of the backside phase mask. This architecture and optimized designs enable diffraction-limited focusing and imaging performance continuously across the near-180$^\circ$ FOV \cite{shalaginov2020single}.

The phase profile of the metasurface will be derived by assuming stigmatic focusing for a pencil of parallel rays incident on the aperture from all directions across the 180$^\circ$ FOV. In the WFOV lens configuration depicted in Fig. \ref{fig:1}b, the phase profile of the metasurface is given by a function $\phi(s)$, where $s$ denotes the radial position from the lens center. Here we consider two parallel rays separated by a small spacing $\Delta s$ both focused by the metasurface to the same point on the image plane. The AOI of the rays inside the substrate is labeled as $\theta$. The stigmatic focusing condition specifies that their propagation path length difference must be exactly compensated by the metasurface, which yields:

\begin{equation}
\Delta s\cdot nsin\theta+\Delta\phi\frac{\lambda}{2\pi}+\left(\frac{\partial}{\partial s} \sqrt{(s-d)^2+f^2}\right)\Delta s=0.
\label{eq:1}
\end{equation}

Here $n$ is the refractive index of the substrate, $\lambda$ is the free-space wavelength and $\Delta\phi$ gives the phase difference the metasurface imparts on the two rays. All other variables are defined following Fig. \ref{fig:1}b. The first term corresponds to the phase difference accumulated at the aperture side, the second term is the one given by metasurface, the third term comes from the difference between the two converging rays separated by distance $\Delta s$ from metasurface to the focal spot. Integration of $\phi$ in Eq. \ref{eq:1} with respect to $s$ reveals the phase profile of the metasurface:

\begin{equation}
\phi(s)=-\frac{2\pi}{\lambda}\int_0^s\left(\frac{ns}{\sqrt{s^2+L^2}}+\frac{s-d}{\sqrt{f^2+(s-d)^2}}\right)ds.
\label{eq:2}
\end{equation}

The only unknown variable in Eq. \ref{eq:1} is $d$, the image height, which is a function of the AOI of the light ray. To determine $d$, we now consider the configuration in Fig. \ref{fig:1}c, where two pencils of parallel rays with slightly different AOIs $\theta$ and $\theta+\Delta\theta$ impinge on the same metasurface area. The two pencils of rays are focused on two different spots on the image plane with image heights of $d$ and $d + \Delta d$, respectively. For the rays with AOI~=~$\theta$, it follows Eq. \ref{eq:1}. Similarly, for the rays with AOI~=~$\theta+\Delta\theta$, the condition becomes:

\begin{equation}
\Delta s\cdot nsin(\theta+\Delta\theta)+\Delta\phi\frac{\lambda}{2\pi}+\left(\frac{\partial}{\partial s}\sqrt{(s-(d+\Delta d))^2+f^2\ }\right)\Delta s=0
\label{eq:3}
\end{equation}

Since the two pencils of rays share the same metasurface area, $\Delta\phi$ is the same for Eqs. \ref{eq:1} and \ref{eq:3} assuming that the angular dependence of meta-atom phase delay is weak, an assumption that is in general satisfied for waveguide-type and resonator-type meta-atoms, which are commonly employed in meta-optics. In the case of strong angular dependence, the second term of Eq. \ref{eq:3} should be modified to include the dependence of AOI, and the phase profile can be similarly derived as follows.  (3) - (1) yields an equation relating $d$ to $\theta$:

\begin{equation}
ncos\theta\Delta\theta+\frac{\partial}{\partial d} \left(\frac{s-d}{\sqrt{(s-d)^2+f^2} }\right)\Delta d=0
\label{eq:4}
\end{equation}

The AOI from free space $\alpha$ is related to $\theta$ via the Snell’s law $sin\alpha=n sin\theta$, and hence Eq. \ref{eq:4} translates to:

\begin{equation}
\Delta d=\left[\left(\frac{Lsin\alpha}{\sqrt{n^2-sin^2\alpha\ }}-d\right)^2+f^2\ \right]^{\frac{3}{2}} \frac{cos\alpha}{f^2}\Delta\alpha
\label{eq:5}
\end{equation}

Substituting Eq. \ref{eq:5} into Eq. \ref{eq:2} leads to the integral form of the target phase profile.

The derivation is generic and applicable to different wavelengths, substrate materials, and meta-atom or diffractive element designs. It can also be extended to cases with multiple substrate layers (with thickness of the $i$-th given by $L_i$). In this case, the new expression of $s=\sum_i L_itan\theta_i$ can be substituted into Eq. \ref{fig:1} and the rest of the analytical formalism remains similar. This is a useful architectural variant which not only opens a larger design space but also allows incorporation of an air gap in between solid substrates to reduce weight or a solid spacer to facilitate fabrication and assembly processes.

\section{Parameter space optimization}

The main assumption in this analytical formalism is that $\Delta s$ is an infinitesimal quantity, which suggests that the ideal stigmatic focusing condition is only rigorously satisfied in the “small aperture” limit. This is intuitive since larger aperture size leads to more spatial overlap of the pencils of rays with different AOIs, which tends to degrade the focusing performance. Next we consider this finite aperture size effect and derive the condition that yields the optimal performance.

When $\Delta s$ is not an infinitesimal quantity, the optical path length difference $\Delta P$ between the two rays in Fig. \ref{fig:1}b can be derived in a manner similar to Eq. \ref{eq:1}:

\begin{equation}
\begin{split}
\Delta P=\Delta s\cdot nsin\theta+\left[\phi\left(s+\Delta s\right)-\phi\left(s\right)\right]\left(\frac{\lambda}{2\pi}\right)\\
+\sqrt{\left(s+\Delta s-d\right)^2+f^2}-\sqrt{\left(s-d\right)^2+f^2}
\label{eq:6}
\end{split}
\end{equation}

To ensure sharp focusing, $\Delta P$ must be minimized. Using Eq. \ref{eq:2} and Eq. \ref{eq:5} and note that $d$ is a function of $s$, we compute the first three orders of derivatives of $\Delta P$ with respect to $\Delta s$ (the detailed derivation process is included in the Appendix A):

\begin{equation}
\frac{d(\Delta P)}{d(\Delta s)}=0
\label{eq:7}
\end{equation}

\begin{equation}
\frac{d^2(\Delta P)}{d(\Delta s)^2}=0
\label{eq:8}
\end{equation}

\begin{equation}
\frac{d^3(\Delta P)}{d(\Delta s)^3}=-\frac{3nL^2(s-d)}{\left(f^2+\left(s-d\right)^2\right)\left(L^2+s^2\right)^{\frac{3}{2}}}
\label{eq:9}
\end{equation}

Denoting aperture diameter as $D$, we compute RMS wavefront error $\sigma$ across the aperture using the derivatives to characterize aberration when $D$ < $f$:

\begin{equation}
\sigma\approx\frac{3nL^2D^3|s-d|}{160\left(f^2+\left(s-d\right)^2\right)(L^2+s^2)^{\frac{3}{2}}}
\label{eq:10}
\end{equation}

This expression explicitly reveals the dependence of lens performance on configuration parameters including focal length, aperture size, substrate thickness and refractive index of substrate. To achieve better performance, one can in general increase the f-number (aka decreasing numerical aperture NA), increase the substrate thickness, and/or reduce the refractive index of substrate. An alternative strategy is to minimize the term $|s-d|$, which implies that a telecentric configuration is conducive to enhanced focusing quality. We further note that this term is dependent on $n$, $L$ and $f$ according to Eq. \ref{eq:5}, which constrains these parameters and explains for example the existence of an optimal substrate thickness for best focusing performance. We want to emphasize that rigorously speaking, our lens structure is not telecentric since the exact condition $s=d$ is inconsistent with Eq. \ref{eq:4} and Eq. \ref{eq:5}. When $\left|s-d\right|$ is much smaller than $f$ and $L$, the aberration becomes dominated by the fourth order derivative and the RMS wavefront error $\sigma$ is (the detailed derivation process is included in the Appendix A): 

\begin{equation}
\sigma\approx\frac{nL^2D^4}{192\sqrt{5}f^2\left(L^2+s^2\right)^\frac{3}{2}}\left|\frac{nL^2f}{\left(L^2+s^2\right)^\frac{3}{2}}-2\right|
\label{eq:11}
\end{equation}

The equation reveals a similar dependence of lens performance on design parameters.

\section{Performance evalution}

\begin{figure*}[htbp]
\centering
\includegraphics[width=\linewidth]{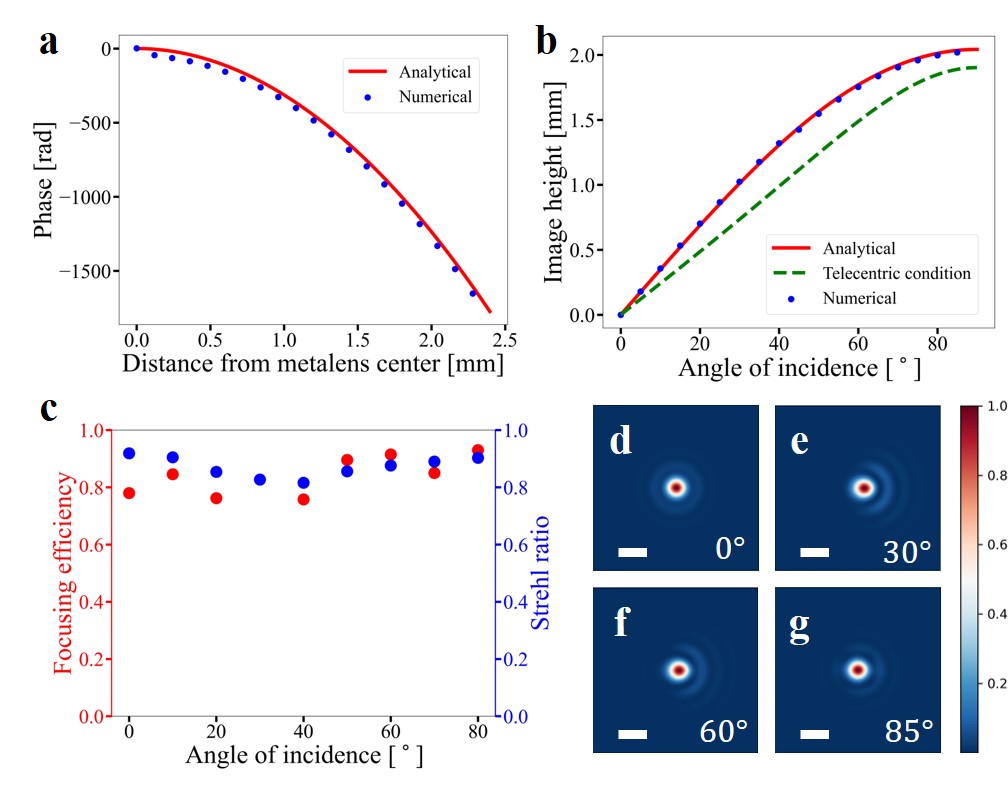}
\caption{Calculated performance of an ideal WFOV lens. (a) Lens phase profile retrieved from analytical and numerical solutions. (b) Image heights with different AOIs from analytical and numerical solutions. The green dashed line represents the telecentric condition which corresponds to $d=s=\frac{Lsin\alpha}{\sqrt{n^2-sin^2\alpha}}$. (c) Focusing efficiency and Strehl ratio for different AOIs. (d)-(g) Normalized intensity profiles at image plane with different AOIs (scale bars are 20 $\mu$m).}
\label{fig:2}
\end{figure*}

We show in the following that the design maintains diffraction-limited performance over the entire hemispherical FOV up to a moderate NA of \textasciitilde 0.25 (corresponding to f/1.9), and that the analytical solution is consistent with numerically optimized designs by considering an exemplary WFOV metalens design operating at 5 $\mu$m wavelength. The lens consists of a 1 mm diameter circular aperture on the front side and a 5 mm diameter circular metasurface on the back side of a 2 mm thick BaF$_2$ substrate ($n$ = 1.45). The effective focal length (spacing between the metasurface and the image plane) is set to 2 mm, corresponding to a NA of 0.24. The analytically derived radial phase profile $\phi$ and image height $d$ are presented in Figs. \ref{fig:2}a and \ref{fig:2}b. As a comparison, we performed numerical optimization using a direct search algorithm \cite{wang2016chromatic,wang2015new,kim2012design,wang2015optical,wang2014computational} (see Appendix B for details), and the optimized phase profiles are plotted in the same graphs. The results confirm excellent agreement between the two approaches.

We then used Kirchhoff diffraction integral \cite{born1991principle} to evaluate the focusing performance of the lens. Assuming a meta-atom pitch of 4 $\mu$m, the lens focusing efficiency (defined as the fraction of power encircled within an area of a diameter equaling to five times the focal spot full-width-at-half-maximum normalized by the total incident power) and Strehl ratio as a function of AOI from air are shown in Fig. \ref{fig:2}c, and the focal spot profiles at several AOIs are displayed in Figs. \ref{fig:2}d-\ref{fig:2}g. The lens exhibits diffraction-limited focusing performance with Strehl ratios consistently larger than 0.8 and efficiencies higher than 75$\%$ over the entire hemispherical FOV.

The diffraction integral calculations above assume ideal meta-atoms so the metasurface acts as a pure phase mask without imposing intensity modulation and phase error. To make a realistic estimate of the metalens efficiency, next we incorporated actual meta-atom structures and their optical characteristics modeled using full-wave calculations \cite{an2020deep}. The all-dielectric, free-form meta-atoms under consideration are made from 1 $\mu$m thick PbTe film resting on a BaF$_2$ substrate \cite{zhang2018ultra,an2019deep}. Properties of the meta-atoms used in the design are tabulated in the Appendix C. The simulation results are shown in Fig. \ref{fig:3}. The focusing efficiency and Strehl ratio are slightly reduced compared to the results in Fig. \ref{fig:2} (which assumes ideal meta-atoms) due to non-unity efficiency and phase error of the simulated meta-atoms. All factors considered, the lens maintains high efficiencies exceeding 65$\%$ and diffraction-limited imaging performance with Strehl ratios above 0.8 across the entire FOV.

\begin{figure}[htbp]
\centering
\includegraphics[width=\linewidth]{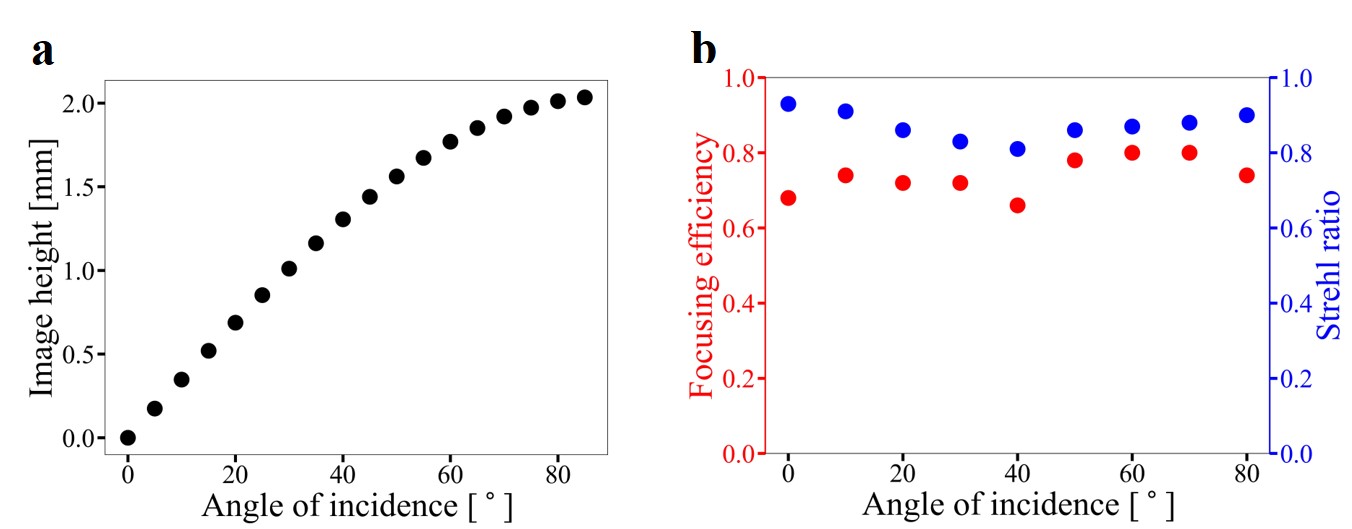}
\caption{Simulated performance of a metalens composed of realistic meta-atoms. (a) Image height, (b) efficiency and Strehl ratio for different AOIs based on full-wave modeled meta-atoms.}
\label{fig:3}
\end{figure}

\section{Design trade-offs investigation}

The analytical formalism allows computationally efficient design of WFOV flat lenses, especially in cases where ray tracing based numerical optimization cannot be implemented in a reasonable time scale. A comparison of the two design methods are presented in the Appendix E. The analytical solution also elucidate the design trade-offs. For a given wavelength and substrate refractive index, the WFOV lens design is fully defined by three independent parameters: aperture size, substrate thickness, and focal length. In the following, we investigate the effect of varying aperture size, substrate thickness and focal length on focusing performance of the lens for a substrate index $n$ = 1.45 and a wavelength $\lambda$ = 5 $\mu$m. The conclusions can be readily generalized to an arbitrary wavelength as the underlying Maxwell’s equations are scale-invariant.

\begin{figure*}[htbp]
\centering
\includegraphics[width=\linewidth]{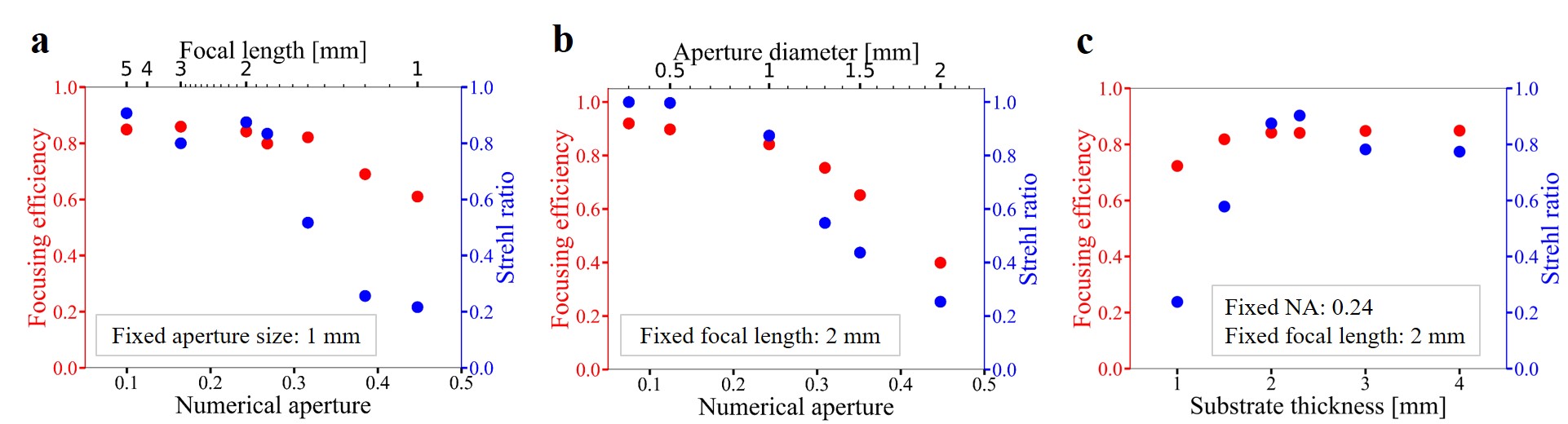}
\caption{(a)-(b) Effects of NA on efficiency and Strehl ratio averaged over the entire near-180$^\circ$ FOV by changing (a) focal length and (b) aperture size. (c) Effects of substrate thickness on averaged efficiency and Strehl ratio.}
\label{fig:4}
\end{figure*}

Figs. \ref{fig:4}a and \ref{fig:4}b plot the focusing efficiency and Strehl ratio values (both averaged over the entire near-180$^\circ$ FOV) for WFOV flat lenses with varying NAs. In Fig. \ref{fig:4}a, the lens aperture diameter is fixed to 1 mm, the substrate thickness is 2 mm, and the focal length is varied to obtain different NAs. In Fig. \ref{fig:4}b, the focal length is set to 2 mm, the substrate thickness is 2 mm, and the aperture diameter is varied. Shorter focal length requires more abrupt change of optical phase whereas spatial overlap between beams with different AOIs increases with larger aperture size, both of which negatively impact the focusing quality. Consequently, both efficiency and Strehl ratio decrease with increasing NA. Fig.~\ref{fig:4}c depicts the impact of varying the substrate thickness. Increasing substrate thickness leads to lower spatial overlap between beams with different AOIs, thereby improving the focusing quality albeit at the expense of larger device footprint. Notably, when the substrate thickness exceeds 2 mm, the design significantly deviates from the telecentric configuration, resulting in lower Strehl ratios as shown in Fig. \ref{fig:4}c. All these results are in accordance with Eq. \ref{eq:10}.

\section{Conclusion}

In summary, we derived an analytical design approach for flat (metasurface or diffractive) fisheye lenses capable of imaging over near-180$^\circ$ FOV. We demonstrate that lenses designed using this scheme can achieve nearly diffraction-limited performance across the entire FOV while maintaining high focusing efficiencies above 65$\%$. This design approach not only sheds light on the key design trade-offs of the WFOV lens, but is also poised to supersede the traditional iterative design scheme and significantly expedite deployment of the WFOV lens technology in diverse applications ranging from 3-D sensing to biomedical imaging.

\begin{appendices}

\section{Derivation of high order derivatives of optical path length difference}

In the following, we present detailed derivation of Eqs. 7-9 in the main text.

\begin{equation}
\Delta P=\Delta s\cdot nsin\theta+\left[\phi(s+\Delta s)-\phi(s)\right]\left(\frac{\lambda}{2\pi}\right)+\sqrt{(s+\Delta s-d)^2+f^2}-\sqrt{(s-d)^2+f^2}
\label{eq:s1}
\end{equation}

\begin{equation}
\frac{\partial(\Delta P)}{\partial(\Delta s)}=nsin\theta+\frac{\partial\phi(s+\Delta s)}{\partial(\Delta s)}\left(\frac{\lambda}{2\pi}\right)+\frac{\partial}{\partial(\Delta s)}\sqrt{(s+\Delta s-d)^2+f^2}
\label{eq:s2}
\end{equation}

The derivative of phase profile can be obtained from Eq. 2 in the main text.

\begin{equation}
\frac{\partial\phi(s+\Delta s)}{\partial(\Delta s)}\left(\frac{\lambda}{2\pi}\right)=-\left[\frac{n(s+\Delta s)}{\sqrt{(s+\Delta s)^2+L^2}}+\frac{s+\Delta s-d(s+\Delta s)}{\sqrt{f^2+(s+\Delta s-d(s+\Delta s))^2}}\right]
\label{eq:s3}
\end{equation}

Substituting Eq. \ref{eq:s3} into Eq. \ref{eq:s2}.

\begin{align}
\frac{\partial(\Delta P)}{\partial(\Delta s)}=&nsin\theta-\frac{n(s+\Delta s)}{\sqrt{(s+\Delta s)^2+L^2}}-\frac{s+\Delta s-d(s+\Delta s)}{\sqrt{f^2+(s+\Delta s-d(s+\Delta s))^2}}\notag\\
&+\frac{\partial}{\partial(\Delta s)}\sqrt{(s+\Delta s-d)^2+f^2}
\label{eq:s4}
\end{align}

Denoting $f(\Delta s)=\sqrt{(s+\Delta s)^2+L^2}$, $g(\Delta s)=\sqrt{f^2+(s+\Delta s-d(s+\Delta s))^2}$,\\
$h(\Delta s)=\sqrt{(s+\Delta s-d)^2+f^2}$, the expression can be simplified as:

\begin{equation}
\frac{\partial(\Delta P)}{\partial(\Delta s)}=nsin\theta-\frac{n(s+\Delta s)}{f(\Delta s)}-\frac{\left(s+\Delta s-d\left(s+\Delta s\right)\right)}{g(\Delta s)}+\frac{\partial h(\Delta s)}{\partial(\Delta s)}
\label{eq:s5}
\end{equation}

The first order derivative of $d$ can be derived from Eq. 4 in the main text:

\begin{equation}
\frac{\partial d(s+\Delta s)}{\partial (\Delta s)}=\frac{nL^2g^3(\Delta s)}{f^2f^3(\Delta s)}
\label{eq:s6}
\end{equation}

The derivatives of $f$, $h$ and $g$ are as follows:

\begin{equation}
\frac{\partial f(\Delta s)}{\partial(\Delta s)}=\frac{s+\Delta s}{f(\Delta s)}
\label{eq:s7}
\end{equation}

\begin{equation}
\frac{\partial h(\Delta s)}{\partial \Delta s}=\frac{s+\Delta s-d}{h(\Delta s)}
\label{eq:s8}
\end{equation}

\begin{equation}
\frac{\partial g(\Delta s)}{\partial(\Delta s)}=\frac{s+\Delta s-d(s+\Delta s)}{g(\Delta s)}\left(1-\frac{\partial{d(s+\Delta s)}}{{\partial \Delta s}}\right)
\label{eq:s9}
\end{equation}

The second and third order derivatives can be obtained from Eq. \ref{eq:s5} - \ref{eq:s9}.

\begin{align}
\frac{\partial^2(\Delta P)}{\partial(\Delta s)^2}&=-\frac{nL^2}{f^3(\Delta s)}-\frac{f^2}{g^3(\Delta s)}\left(1-\frac{\partial d(s+\Delta s)}{\partial(\Delta s)}\right)+\frac{\partial^2h(\Delta s)}{\partial(\Delta s)^2}\notag\\
&=-\frac{f^2}{g^3(\Delta s)}+\frac{f^2}{h^3(\Delta s)}
\label{eq:s10}
\end{align}

\begin{equation}
\frac{\partial^3(\Delta P)}{\partial(\Delta s)^3}=\frac{3(s+\Delta s-d(s+\Delta s))}{g^2(\Delta s)}\left(\frac{f^2}{g^3(\Delta s)}-\frac{nL^2}{f^3(\Delta s)}\right)-\frac{3f^2(s+\Delta s-d)}{h^5(\Delta s)}
\label{eq:s11}
\end{equation}

Taking $\Delta s\to 0$ which corresponds to the center of the aperture:

\begin{equation}
\beta_1=\frac{\partial(\Delta P)}{\partial(\Delta s)}_{\Delta s\rightarrow 0}=nsin\theta-\frac{ns}{\sqrt{s^2+f^2}}-\frac{s-d}{\sqrt{f^2+(s-d)^2}}+\frac{s-d}{\sqrt{(s-d)^2+f^2}}=0
\label{eq:s12}
\end{equation}

\begin{equation}
\beta_2=\frac{\partial^2(\Delta P)}{\partial(\Delta s)^2}_{\Delta s\rightarrow 0}=-\frac{f^2}{\left [ f^2+(s-d)^2\right]^\frac{3}{2}}+\frac{f^2}{\left [ f^2+(s-d)^2\right]^\frac{3}{2}}=0
\label{eq:s13}
\end{equation}

\begin{equation}
\beta_3={\frac{\partial^3(\Delta P)}{\partial(\Delta s)^3}}_{\Delta s\to 0}=-\frac{3nL^2(s-d)}{(f^2+(s-d)^2)(L^2+s^2)^{\frac{3}{2}}}
\label{eq:s14}
\end{equation}

The design space can be divided into three regimes. When the aperture diameter is larger than the focal length ($D>f$), which corresponds to numerical apertures (NAs) greater than 0.45, the solution based on Taylor expansion fails to accurately account for the optical length difference (OPD). This regime is not the main focus of our discussion since the imaging quality of wide FOV lenses degrades with very large NA. When $D<f$ and $\left|s-d\right|$ is comparable to $f$ and $L$, OPL is dominated by $\beta_3$:

\begin{equation}
\Delta P\approx\frac{1}{6}\beta_3(\Delta s)^3
\label{eq:s15}
\end{equation}

Denoting $\rho=\frac{2\Delta s}{D}$, the RMS wavefront error $\sigma$ can be computed as follow:

\begin{equation}
\sigma^2=2\int_0^1\left[\Delta P(\rho)-\overline{\Delta P}\right]^2\rho d\rho
\label{eq:s16}
\end{equation}

\begin{equation}
\sigma=\frac{1}{160}\left|\beta_3\right|D^3
\label{eq:s17}
\end{equation}

\noindent which corresponds to Eq. 10 in the main text. The last case is $D<f$ and $\left|s-d\right|\ll f,L$, which corresponds to a structure close to the telecentric configuration and yielding improved performance. In this case, OPD is dominated by fourth order derivative which can be derived from Eq. \ref{eq:s11} with all terms containing $\left|s-d\right|$ omitted:

\begin{equation}
\beta_4=\frac{\partial^4(\Delta P)}{\partial(\Delta s)^4}\approx\frac{3nL^2}{f^2(L^2+s^2)^\frac{3}{2}}\left(\frac{nL^2f}{\left(L^2+s^2\right)^\frac{3}{2}}-2\right)
\label{eq:s18}
\end{equation}

\begin{equation}
\Delta P\approx\frac{1}{24}\beta_4(\Delta s)^4
\label{eq:s19}
\end{equation}

\begin{equation}
\sigma=\frac{1}{576\sqrt{5}}\left|\beta_4\right|D^4
\label{eq:s20}
\end{equation}

\noindent which leads to Eq. 11 in the main text.

\section{Numerical optimization of phase mask}

\begin{figure}[htbp]
\centering
\includegraphics[width=\linewidth]{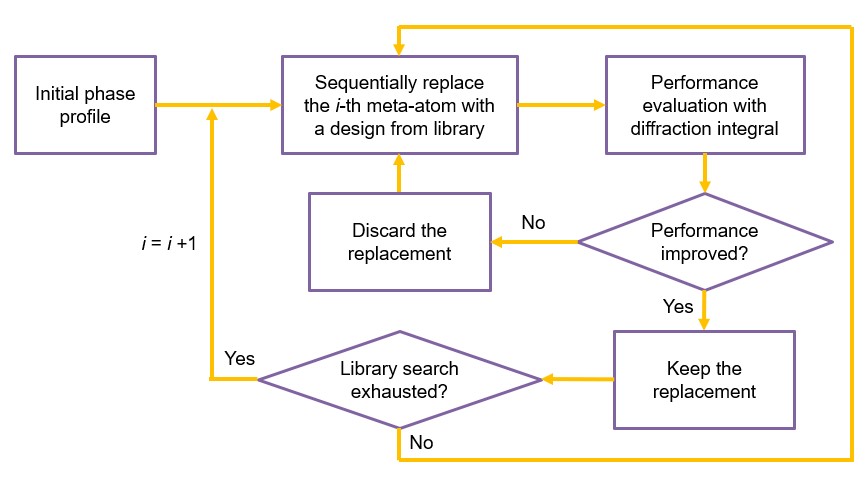}
\caption{Flow chart of the direct search algorithm used for numerical verification.}
\label{fig:s1}
\end{figure}

Numerical optimization was used to validate the analytical solution. It employs a direct search algorithm to find the optimum phase profile of the lens and the image height as a function of angles of incidence (AOIs). The flow chart illustrating the direct search algorithm is shown in Fig. \ref{fig:s1}. For this wide field-of-view (WFOV) optical lens, the figure-of-merit (FOM) is defined as the sum of peak intensities at the focal spots of all the AOIs sampled:

\begin{equation}
FOM=\sum_i I_{AOI(i)}
\label{eq:s21}
\end{equation}

The analytical solution is first applied to estimate the image height. Then, the algorithm searches the image plane within a 40 $\mu$m range to find the intensity peak, which is then defined as the numerically derived image height and is used to compute the FOM.

The direct search algorithm starts with a randomly generated phase profile of the lens and compute the initial FOM. Then, it traverses all meta-atom positions with 4 $\mu$m sampling spacing, which is the pitch of meta-atoms. For every position, different phase delays are sequentially tested and the one with largest FOM is selected. After traversing the entire metasurface, the final phase mask is adopted as the optimum one, which is shown in Fig \ref{fig:2}a of the main text.

\section{Meta-atom design}

\begin{figure}[htbp]
\centering
\includegraphics[width=\linewidth]{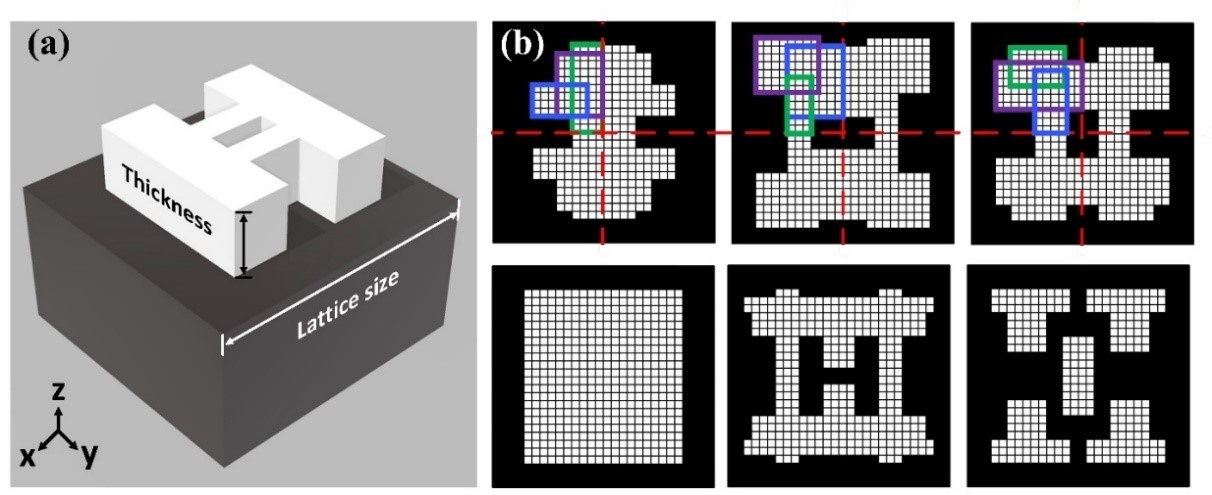}
\caption{Examples of generated meta-atom structures. (a) 3-D view. (b) Several examples of generated 2-D patterns. Rectangles outlined in different colors represent randomly generated high-index "Needles".}
\label{fig:s2}
\end{figure}

Meta-atoms are composed of 1 $\mu$m thick dielectric PbTe resting on a BaF$_2$ substrate with a pitch of 4 $\mu$m. Sample meta-atom structures are shown in Fig. \ref{fig:s2}. The 2-D pattern of each meta-atom was generated with a "Needle Drop" approach. Several rectangular bars, with a minimum generative resolution of 1 pixel, were randomly generated and placed together with a square canvas (64 $\times$ 64 pixels) to form random patterns. To minimize inter-cell coupling, a minimum spacing of 8 pixels was applied between adjacent meta-atoms.

\begin{figure}[htbp]
\centering
\includegraphics[width=\linewidth]{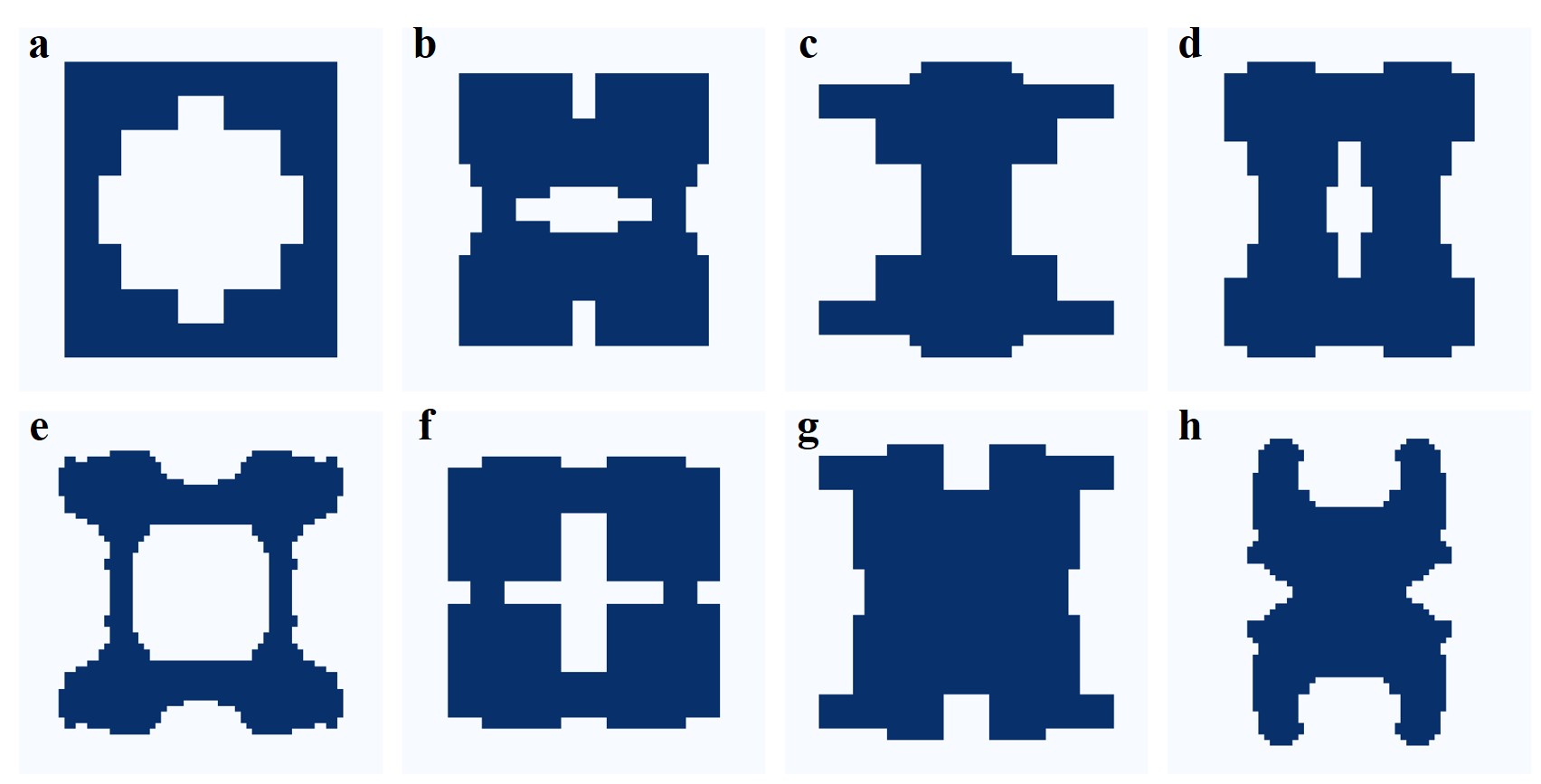}
\caption{Selected 8 meta-atom structures.}
\label{fig:s3}
\end{figure}

\begin{table}[htbp]
\centering
\caption{\bf Meta-atom phase delays and transmittance values}
\begin{tabular}{ccccccccc}
\hline
Index & a & b & c & d & e & f & g & h \\
\hline
Phase delays [$^\circ$] & 0.7 & 45.0 & 90.2 & 135.0 & 180.1 & 224.8 & 269.9 & 315.3 \\
Transmittance & 0.84 & 0.93 & 0.93 & 0.95 & 0.89 & 0.90 & 0.92 & 0.94 \\
\hline
\end{tabular}
  \label{tab:s1}
\end{table}

Full-wave electromagnetic simulations were performed with TE polarization (electric field always parallel to the horizontal direction in the figures) to obtain the phase delays of these generated structures. Then, the phase delays were discretized into 8 groups with $\frac{\pi}{4}$ spacing, and the optimum structures were chosen according to a figure-of-merit which aims to minimize phase deviations and maximize transmission amplitudes \cite{shalaginov2021reconfigurable}. The 8 selected meta-atom structures are shown in Fig. \ref{fig:s3}. Their phase delays and transmittance values are summarized in Table \ref{tab:s1}. These meta-atoms were utilized in the simulation in Fig. 3 of the main text.

\section{Diffraction integral characterization of lens performance}

\begin{figure}[htbp]
\centering
\includegraphics[width=\linewidth]{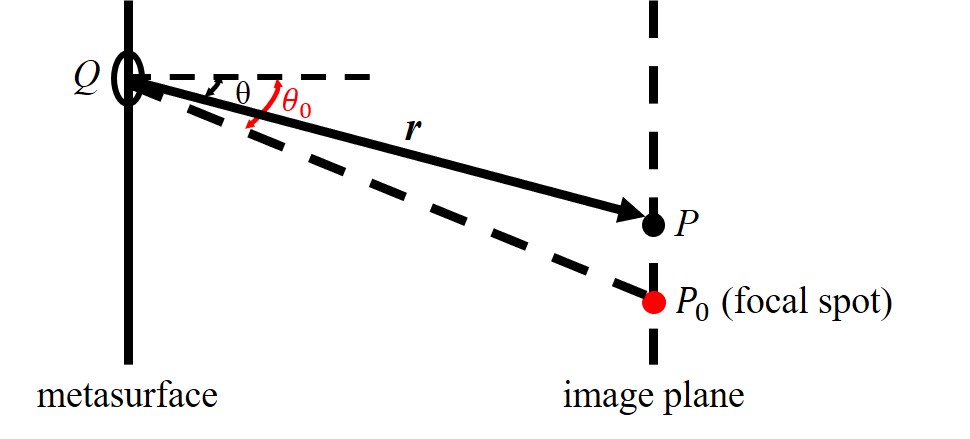}
\caption{Illustration of Kirchhoff diffraction integral method.}
\label{fig:s4}
\end{figure}

The Kirchhoff diffraction integral was utilized to evaluate the focusing performance of the lens. The configuration is depicted in Fig. \ref{fig:s4}. The diffraction integral is expressed as:

\begin{equation}
\widetilde{U}(P)=\frac{-i}{\lambda}\cdot \Lambda^2 \sum_N \frac{1}{2} (cos\theta_0+cos\theta)\ \widetilde{U_0}(Q)\cdot\frac{1}{r}\cdot e^{ikr}
\label{eq:s22}
\end{equation}

Here $\lambda$ is the wavelength, $\Lambda=4\ \mu m$ is the pitch of meta-atoms, $ \widetilde{U_0}(Q)$ is the complex amplitude incident on the metasurface, and $\widetilde{U}(P)$ is the complex amplitude on the image plane. Since most of the incident waves converge near the focal spot, the first angular term $\theta_0$ in the integration, which is the direction of outlet wave, is taken as the angle between surface normal and focal spot direction as an approximation. The approximation is accurate for lenses with high Strehl ratios, although we have numerically verified that this condition yields < 5 percent error even in the case of a low Strehl ratio (0.3). The summation includes all meta-atoms on the surface.

To evaluate the lens efficiency and Strehl ratio, Kirchhoff diffraction integral was first utilized to generate intensity profile at the image plane at different AOIs, followed by calculation of the full-width-at-half-maximum (FWHM) of the focal spots. The focusing efficiency is defined as the fraction of power encircled within an area of a diameter equaling to five times the focal spot FWHM normalized by the total incident power, and the Strehl ratio is defined as the ratio of the simulated peak intensity to that of an ideal aberration-free lens of the same focusing efficiency. These results are shown in Figs. 2-4 of the main text.

\section{Comparison between the analytical formalism and ray tracing based optimization}

\begin{figure}[htbp]
\centering
\includegraphics[width=8cm]{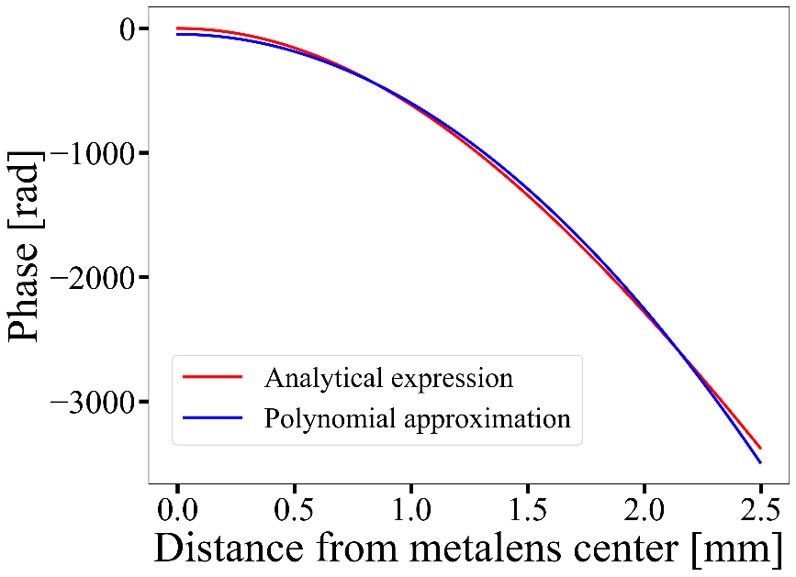}
\caption{Phase profile comparison between the analytical expression and a second order polynomial (quadratic) approximation with an aperture size of 0.5 mm and a focal length of 1 mm.}
\label{fig:s5}
\end{figure}

\begin{figure}[htbp]
\centering
\includegraphics[width=\linewidth]{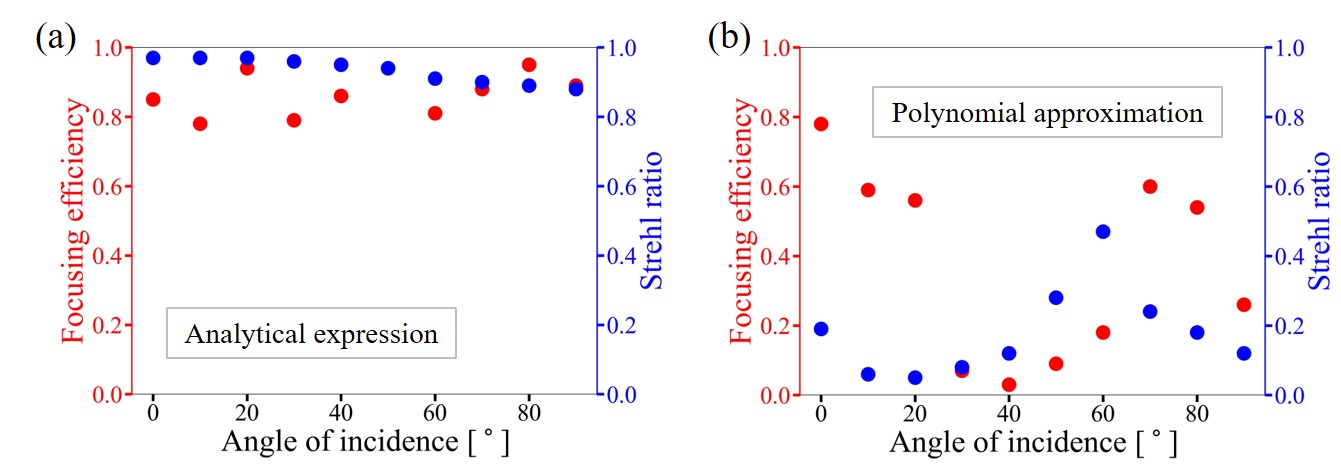}
\caption{Focusing efficiency and Strehl ratio with different angles of incidence using the phase profiles derived from (a) the analytical expression and (b) the second order polynomial (quadratic) approximation.}
\label{fig:s6}
\end{figure}

The phase profile with the design example in the main text happens to be very close to a second order polynomial, which means that ray tracing can also give similar result by using only one power term. This is, coincidentally, also the design reported by Engelberg \emph{et al.} \cite{engelberg2020near}. This is a coincidence because of the chosen structure parameters and does not characterize the general scenario. For example, if both aperture size and focal length are reduced by half (aperture size 0.5 mm and focal length 1 mm), the phase profiles of the analytical expression and the second order polynomial approximation are compared in Fig. \ref{fig:s5}. It can be seen that there is a notable difference. The efficiencies and Strehl ratios with different AOIs using these two phases are compared in Fig. \ref{fig:s6}, which show very different results. Therefore, brute-force ray tracing optimization requires simulation of higher order terms to achieve comparable performance with the analytical solution in most cases, which demands longer computation time considering that all incident angles need to be modeled. Moreover, the optimization process becomes far more complex when the substrate is not a uniform monolith (e.g. with an air gap or a solid spacer), a case which our analytical formalism also readily covers. Finally, when considering a broadband WFOV lens, the analytical model can expediently generate the optimum phase profiles for different wavelengths. In comparison, since each wavelength requires separate optimization of the phase profile, the computation load becomes excessively large for a brute-force optimization scheme.

\end{appendices}

\section{Funding}
Defense Advanced Research Projects Agency Defense Sciences Office (DSO) Program: EXTREME Optics and Imaging (EXTREME) under Agreement No. HR00111720029; MIT Skoltech Seed Fund Program; MIT Deshpande Center for Technological Innovation.

\section{Disclosures}
The authors declare no conflicts of interest.

\section{Data Availability Statement}
The simulation data are available upon reasonable request.

\bibliographystyle{apalike-refs}
\bibliography{reference}

\end{document}